\documentclass[12pt]{article}
\usepackage{latexsym,a4}

\newcommand{\bdm}{\begin{displaymath}}
\newcommand{\edm}{\end{displaymath}}
\newcommand{\no}{\nonumber \\}

\newcommand{\be}{\begin{equation}}
\newcommand{\ee}{\end{equation}}
\newcommand{\bea}{\begin{eqnarray}}
\newcommand{\eea}{\end{eqnarray}}

\newcommand{\co}{\; \; ,}

\newcommand{\epe}{\varepsilon'/\varepsilon}

\begin{document}
\begin{titlepage}

\begin{flushright}ZU-TH 8/01 \end{flushright}
\vspace{3cm}
\begin{center}{\Large\bf A note on the dispersive treatment of $K \to \pi
    \pi$ with the kaon off-shell}

\vspace{2 cm}
M.~B\"uchler, G.~Colangelo, J.~Kambor and F.~Orellana

\vspace{1cm}
Institute for Theoretical Physics, University of Z\"urich\\
Winterthurerstr. 190, CH-8057 Z\"urich, Switzerland\\

\vspace{1cm}
February 2001
\end{center}

\vspace{4cm}

\begin{abstract}
It has been recently suggested that it is possible to calculate the effect
of final state interactions in $K \to \pi \pi$ amplitudes by applying
dispersive methods to the amplitude with the kaon off-shell. We critically
reexamine the procedure, and point out the effects of the arbitrariness in
the choice of the off-shell field for the kaon.
\end{abstract}
\end{titlepage}

\noindent {\bf 1.}
In two recent papers \cite{PP}, Pallante and Pich have pointed out that if
one includes the effect of final state interactions (FSI) in the
calculation of the weak $K \to \pi \pi$ matrix elements, one may bring the
Standard Model (SM) calculation of $\epe$ into agreement with the measured
value. In their treatment of the problem they have followed an old
suggestion by Truong \cite{truong}, who showed that final state
interactions produce an effect that goes in the right direction to produce
the $\Delta I = 1/2$ rule. Unfortunately, the size of the effect is too
small to fully explain the rule. On the other hand, in the case of $\epe$,
FSI seem to give just about the right correction to yield the measured
value from the SM calculation.

In this note we critically reexamine the proposed procedure, and point out
the problems that one has to face if one tries to solve the dispersion
relations for the off-shell amplitude. In essence, the main problem with
this approach is that there are infinitely many ways in which one can go
off-shell (see also \cite{suzuki}), while, on the other hand, the on-shell
amplitude is unique: in the language of dispersion relations this means
that different sets of subtraction constants have to lead to the same
result for the on-shell amplitude. Making this procedure work in practice
may be problematic.

This arbitrariness in going off-shell with the kaon has been circumvented
in Ref. \cite{buras}, where a dispersion relation in the mass of the kaon
has been formulated. In this case, however, one lacks a rigorous framework
for discussing and implementing the dispersion relation.  Alternatively,
one can avoid going off-shell with the kaon by allowing the weak
Hamiltonian to carry momentum: the amplitude then becomes a function of
three Mandelstam variables, and the corresponding dispersion relations are
more complicated. Nonetheless, they can be solved numerically without major
difficulties. This framework has an important advantage: that one can use
soft-pion theorems to fix the subtraction constants.  We discuss this
approach in a separate paper \cite{kpp}.

\vskip 0.5cm
\noindent {\bf 2.}
We will denote the generic interpolating field of the kaon with $X^K$. In
what follows we will consider only two choices\footnote{We stress that the
  dynamical fields appearing in the effective Lagrangian cannot be
  meaningfully used to go off-shell.}: 
\be 
A^K_\mu=\bar s \gamma_\mu
\gamma_5 d\; , \; \; \; \; P^K= \bar s \gamma_5 d \; ,
\label{eq:AP}
\ee 
but one could of course combine these in various ways, take derivatives,
etc. -- all would lead to a perfectly well defined off--shell amplitude,
and all give the same on--shell amplitude (modulo overall factors). 
The object that we will consider is the following: 
\be 
G_X(s)={iN_X} \int dx e^{ikx} \langle \pi(p_1) \pi(p_2) {\rm out}| T
\left( {\cal H}_W(0) X^K(x)\right) | 0 \rangle \; , \; \; \; \; s=k^2\; , 
\ee 
$k=p_1+p_2$. $N_X$ is a normalization factor (possibly a Lorentz vector)
which depends on the interpolating field, and which is defined such that
the residue of the pole at $s=M_K^2$ is the same for all possible
interpolating fields:  
\be
G_X(s) = { {\cal A} \over s-M_K^2} + B_X(s) \; \; , 
\ee 
and is the $K \to \pi \pi$ amplitude.

In order to set up a dispersion relation for $G_X(s)$ we need to know its
analytic properties: besides the pole at $s=M_K^2$, it has a cut starting
at $s=4 M_\pi^2$, and is analytic everywhere else. If we assume that one
subtraction constant is sufficient\footnote{The point we want to make does
  not depend on the number of subtractions that are necessary.}, we may
write the following dispersion relation
\be
\label{eq:Gdisprel}
G_X(s) = G_X(s_0)+ {(s-s_0) {\cal A} \over (M_K^2-s_0)(s-M_K^2)} + (s-s_0)
\int_{4 M_\pi^2}^{\infty} ds' {\rm{disc} [G_X(s')] \over (s'-s_0)(s'-s)} \;
\; . 
\ee 
In order to solve the dispersion relation, we neglect the
contributions to the discontinuity coming from inelastic channels, and
assume that the phase of $G_X(s)$ at the upper rim of the cut is given by
the $\pi \pi$ phase shift $\delta(s)$, all the way up to infinity. The
explicit solution of this dispersion relation is a simple modification
(which accounts for the presence of the pole) of the Omn\`es solution
\cite{omnes}: 
\be
\label{eq:Gomnes}
G_X(s)=\left[G_X(s_0)+ {(s-s_0) {\cal A} \over (M_K^2-s_0)(s-M_K^2)
    \Omega(M^2,s_0) } \right] \Omega(s,s_0) \; \; ,
\ee
where $\Omega(s,s_0)$ is the once--subtracted Omn\`es function, defined as:
\be
\Omega(s,s_0) = \exp\left\{ {(s-s_0) \over \pi} \int_{4 M_\pi^2}^{\infty}
  ds' {\delta(s) \over (s'-s_0)(s'-s)} \right\} \; \; .
\ee
Both Eqs. (\ref{eq:Gdisprel}) and (\ref{eq:Gomnes}) show clearly the
obvious fact that the value of the residue is one of the inputs, and cannot
be obtained as output from the dispersion relation for $G_X(s)$.

\vskip 0.5cm
\noindent {\bf 3.}
However, one can consider the function $G_X(s)$ multiplied by $s-M_K^2$:
\be
F_X(s)=(s-M_K^2) G_X(s) \; \; ,
\ee
and apply a dispersion relation to this function. By definition, $F_X(s)$
needs one more subtraction than $G_X(s)$, and has of course no pole at
$s=M_K^2$. A dispersion relation for $F_X(s)$ reads:
\be
F_X(s)=F_X(s_0)+(s-s_0)F_X'(s_0)+ (s-s_0)^2 \int_{4 M_\pi^2}^{\infty} ds'
{\rm{disc} [F_X(s')] \over (s'-s_0)^2(s'-s)} \; \; 
,
\ee
and its solution (within the same approximation as above) is of the general
form due to Omn\`es \cite{omnes} 
\be
\label{eq:Fomnes}
F_X(s)=\left\{F_X(s_0)+(s-s_0)\left[F_X'(s_0)-F_X(s_0)
    \Omega'(s_0,s_0)\right] \right\} \Omega(s,s_0) \; \; ,
\ee
where both $F_X'(s)$ and $\Omega'(s,s_0)$ are first derivatives in $s$.
To get ${\cal A}$ we simply have to evaluate the general solution
(\ref{eq:Fomnes}) at $s=M_K^2$. The result is
\be
\label{eq:Aomnes}
{\cal A}=\left\{F_X(s_0)+(M_K^2-s_0)\left[F_X'(s_0)-F_X(s_0)
    \Omega'(s_0,s_0)\right] \right\} \Omega(M_K^2,s_0) \; \; .
\ee
It is easy to verify that if one substitutes back $F_X(s)=(s-M_K^2)
G_X(s)$, and uses the solution (\ref{eq:Gomnes}), Eq. (\ref{eq:Aomnes})
becomes an identity. In other words, Eq. (\ref{eq:Aomnes}) explicitly shows
that the value of the residue of the pole at $s=M_K^2$ of $G_X(s)$ is
hidden inside the derivative of $F_X(s)$ at $s=s_0$, and gives a recipe for
subtracting out the contribution proportional to $G_X(s_0) \Omega'(s,s_0)$.
If we had a means to calculate both $F_X(s_0)$ and $F_X'(s_0)$ reliably, 
more than the amplitude ${\cal A}$ itself, then we could indeed use
Eq. (\ref{eq:Aomnes}) to obtain the $K \to \pi \pi$ amplitude. As a check
one should verify that the final result does not depend on the choice of
the interpolating field $X^K$. We are not aware of any methods that could
give $F_X(s_0)$ and $F_X'(s_0)$ more reliably than the amplitude itself.
In particular, since below threshold $F_X(s)$ is defined via analytic
continuation, numerical methods cannot directly calculate the subtraction
constants.

\vskip 0.5cm
\noindent {\bf 4.}
To better illustrate the content of Eq. (\ref{eq:Aomnes}), we find it
useful to apply the chiral counting to the subtraction constants that
appear in there. We define the following expansion:
\bea
G_X(s_0)&=&G_X^{(0)}(s_0)+G_X^{(2)}(s_0)+O(p^4) \; \; ,\nonumber \\
{\cal A} &=& {\cal A}^{(2)}+{\cal A}^{(4)}+O(p^6) \; \; ,
\eea
which translates into a chiral expansion for $F_X(s_0)$ and
$F_X'(s_0)$. Notice that because of the different physical dimensions we
have indicated the leading chiral order for $G_X(s_0)$ (${\cal A}$) as
$p^0$ ($p^2$). 

If we use Eq. (\ref{eq:Aomnes}) and fix the subtraction constants at
leading chiral order, we get
\be
\label{eq:A1}
{\cal A} = \left[{\cal A}^{(2)}-G_X^{(0)}(s_0) (M_K^2-s_0)^2
  \Omega'(s_0,s_0) \right] \Omega(M_K^2,s_0) \; \; ,
\ee
a result which shows an unwanted dependence on the choice of the
interpolating field $X^K$. 
To exemplify, we consider the two interpolating fields in (\ref{eq:AP}),
and find: 
\be
G_P^{(0)}\sim 2c_2-{4\over 3}c_5\left(1+{M_\pi^2 \over 2 M_K^2} \right) ~~~
~~~ G_{A}^{(0)} \sim c_2 \co
\ee
where $c_{2,5}$ are coupling constants defined in \cite{EKW}, and where we
have neglected an uninteresting normalization factor.
Numerically, the correction depending on the interpolating field is fairly
sizable:
\bea
{\cal A}_{A}^{(0)} &=& {\cal A}^{(2)}\left[1-
  { (M_K^2-s_0)^2 \over 2 (M_K^2-M_\pi^2)} \Omega'(s_0,s_0) \right] 
\Omega(M_K^2,s_0)\no
&=&{\cal A}^{(2)}\left[1-0.26\right] \Omega(M_K^2,M_\pi^2)
\co 
\eea
where the last equality follows for $s_0=M_\pi^2$. In evaluating the
Omn\`es function we have cut off the dispersive integral at 1 GeV.
We cannot evaluate numerically the case with the pseudoscalar interpolating
field, because we do not know $c_5/c_2$. However this simple numerical
exercise shows that the arbitrary correction is numerically relevant for
the final result. 

In this example we have full control over the chiral order of each term. It
is then easy to remove by hand the term which carries the dependence on the
interpolating field, the term proportional to $G_X^{(0)}(s_0)$ in
Eq. (\ref{eq:A1}). The end result in \cite{PP} and \cite{truong} can be
viewed as an implementation of this procedure\footnote{G.C. thanks Toni
  Pich for a clarifying discussion on this point.}.
We stress, however, that the corresponding result does not follow from a
rigorous application of dispersion relations, but it is rather a
dispersion-relation inspired method to resum rescattering effects. In
particular, in this manner one cannot resolve the arbitrariness at the
level of finite terms that may be moved at will from the subtraction
polynomial to the exponential in the Omn\`es function. A thorough
discussion of the latter point can be found in Ref. \cite{GM}, where such a
resummation method (baptised there as the ``Modified Omn\`es
representation'') had been implemented in the case of the scalar form
factor of the pion.

It is reassuring to see that if we fix the subtraction constants at
next--to--leading order in the chiral expansion, the arbitrariness shows up
one order higher:
\be
\label{eq:A3}
 \! \! \!{\cal A} \!= \! \left[{\cal A}^{(2)} \!\left( \!1 \!-
    \!\Omega^{(2)}(M_K^2,s_0) \!\right) 
   \!+ \!{\cal A}^{(4)} \!- \!G_X^{(2)}(s_0) (M_K^2-s_0)^2
  \Omega'(s_0,s_0) \! \right] \! \Omega(M_K^2,s_0)  \; ,
\ee
where $\Omega^{(2)}(s,s_0)$ is the contribution of order $p^2$ of the
Omn\`es function:
\be
\Omega^{(2)}(s,s_0) =  {(s-s_0) \over \pi} \int_{4 M_\pi^2}^{\Lambda^2}
  ds' {\delta^{(2)}(s) \over (s'-s_0)(s'-s)} \; \; .
\ee
Extending Eq. (\ref{eq:A3}) to yet higher orders is trivial.

\vskip 0.5cm
\noindent {\bf 5.}
We summarise our main conclusions:
\begin{enumerate}
\item
If one knows $F_X(s_0)$ and $F_X'(s_0)$, and the Omn\`es function, one can
indeed obtain the amplitude ${\cal A}$, using Eq. (\ref{eq:Aomnes}). This
result simply follows from analyticity.
\item 
In practical terms the method works only if one has a means to get
the two subtraction constants more accurately than the physical amplitude
itself -- in our opinion this is the main problem with this approach,
because, as far as we know, no such methods are available at present.
\item
Combining Eq. (\ref{eq:Aomnes}) with the chiral expansion is very
instructive, and shows that the arbitrariness connected with the choice of
the interpolating field for going off--shell always appears one order
higher than the one used to fix the subtraction constants.
\end{enumerate}

\vskip 0.5cm \noindent
{\bf Acknowledgements}~~ It is a pleasure to thank J.~Gasser, G.~Isidori,
E.~Pallante, T.~Pich and D.~Wyler for interesting discussions. 
This work was supported by the Swiss National Science
Foundation, by TMR, BBW-Contract No. 97.0131 and EC-Contract
No. ERBFMRX-CT980169 (EURODA$\Phi$NE).

\end{document}